\documentclass[reprint,superscriptaddress,twocolumn,amsmath,amssymb,aps,prl]{revtex4-2}
\usepackage{graphicx}
\usepackage{amsmath}
\usepackage{amssymb}
\usepackage{braket}
\usepackage{dcolumn}
\usepackage{upgreek}
\usepackage{tikz}
\usepackage[colorlinks=true,linkcolor=blue,citecolor=blue,urlcolor=blue]{hyperref}
\usepackage{natbib} 
\usepackage[normalem]{ulem}
\usepackage{url}  

\begin{document}

\title{Direct Generation of an Array with 78400 Optical Tweezers Using a Single Metasurface}
\author{Yuqing Wang}
\thanks{These authors contributed equally to this work.}
\affiliation{Department of Physics and State Key Laboratory of Low Dimensional Quantum Physics, Tsinghua University, 100084, Beijing, China.}
\author{Yuxuan Liao}
\thanks{These authors contributed equally to this work.}
\affiliation{Department of Electronic Engineering, Tsinghua University, 100084, Beijing, China.}
\author{Tao Zhang}
\author{Ye Tian}
\author{Yujia Wu}
\author{Wenjun Zhang}
\affiliation{Department of Physics and State Key Laboratory of Low Dimensional Quantum Physics, Tsinghua University, 100084, Beijing, China.}
\author{Wei Zhang}
\affiliation{Department of Electronic Engineering, Tsinghua University, 100084, Beijing, China.}
\author{Yidong Huang}
\affiliation{Department of Electronic Engineering, Tsinghua University, 100084, Beijing, China.}
\author{Hui Zhai}
\affiliation{Institute for Advanced Study, Tsinghua University, 100084, Beijing, China.}
\author{Wenlan Chen}
\affiliation{Department of Physics and State Key Laboratory of Low Dimensional Quantum Physics, Tsinghua University, 100084, Beijing, China.}
\author{Xue Feng}
\email{x-feng@tsinghua.edu.cn}
\affiliation{Department of Electronic Engineering, Tsinghua University, 100084, Beijing, China.}
\author{Zhongchi Zhang}
\email{zhang-zc20@mails.tsinghua.edu.cn}
\affiliation{Department of Physics and State Key Laboratory of Low Dimensional Quantum Physics, Tsinghua University, 100084, Beijing, China.}
\date{\today}

\begin{abstract}

Scalability remains a major challenge in building practical fault-tolerant quantum computers. Currently, the largest number of qubits achieved across leading quantum platforms ranges from hundreds to thousands. In atom arrays, scalability is primarily constrained by the capacity to generate large numbers of optical tweezers, and conventional techniques using acousto-optic deflectors or spatial light modulators struggle to produce arrays much beyond $\sim 10,000$ tweezers. Moreover, these methods require additional microscope objectives to focus the light into micrometer-sized spots, which further complicates system integration and scalability. Here, we demonstrate the experimental generation of an optical tweezer array containing $280\times 280$ spots using a metasurface, nearly an order of magnitude more than most existing systems. The metasurface leverages a large number of subwavelength phase-control pixels to engineer the wavefront of the incident light, enabling both large-scale tweezer generation and direct focusing into micron-scale spots without the need for a microscope. This result shifts the scalability bottleneck for atom arrays from the tweezer generation hardware to the available laser power. Furthermore, the array shows excellent intensity uniformity exceeding $90\%$, making it suitable for homogeneous single-atom loading and paving the way for trapping arrays of more than $10,000$ atoms in the near future.

\end{abstract}

\maketitle

The pursuit of a universal, fault-tolerant, and practical quantum computer stands as one of the most ambitious goals in modern quantum science and technology~\cite{nielsen2010quantum}. A central challenge in this endeavor is achieving scalability in the number of qubits. This becomes especially critical in the fault-tolerant era, where quantum error correction trades qubit resources for increased computational accuracy by organizing multiple physical qubits into a single logical qubit. Among various platforms, atom arrays have demonstrated significant promise owing to the identical particle nature of cold atoms, and scalability in this architecture is constrained only by the scale and capability of the optical tweezers used to trap individual atoms~\cite{endres2016atom, barredo2018synthetic, ebadi2021quantum, chiu2025continuous, lin2025ai, manetsch2025tweezer}.

\begin{figure}
    \begin{center}  
    \includegraphics[width=0.8\columnwidth]{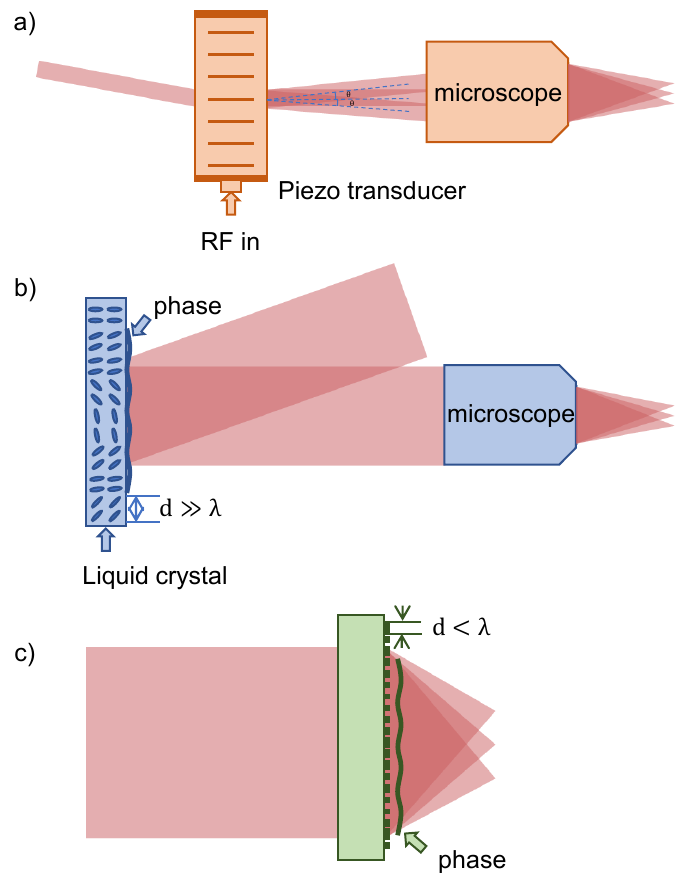}
    \caption{Comparison of technologies for generating optical tweezers. (a) A pair of AODs uses radio-frequency signals to produce acoustic waves that diffract the incident light into a two-dimensional array. (b) An SLM generates arbitrary patterns by encoding a phase hologram on a pixelated liquid-crystal array. Its micron-scale pixel size, however, limits scalability. Both (a) and (b) require a microscope objective to focus the light spots into micrometer-scale optical tweezers. (c) An optical metasurface directly shapes the wavefront of light via a dense array of subwavelength nanostructures, allowing the generation of multiple tightly focused optical tweezers in a compact, lensless configuration. } 
    \label{comparing}
        \end{center}
\end{figure}

Currently, the optical tweezers in atom array platforms are commonly generated using acousto-optic deflectors (AODs) or spatial light modulators (SLMs)~\cite{endres2016atom,bernien2017probing,barredo2018synthetic,Madjarov2020,ebadi2021quantum, cong2022hardware, Burgers2022, deist2022mid, bluvstein2022quantum, evered2023high, Ma2023,anand2024dual,zhang2024scaled, finkelstein2024universal, muniz2025high,manetsch2025tweezer, tsai2025benchmarking, bluvstein2025architectural, chiu2025continuous,lin2025ai, zhang2025observation}, which typically achieve arrays on the order of a few thousand tweezers and barely reach tens of thousands, due to several serious technical challenges in AODs and the finite pixel count and limited diffraction efficiency of SLMs. To date, the highest reported number with these approaches is approximately 12,000 tweezers, used to trap 6,100 atoms by combining two SLMs~\cite{manetsch2025tweezer}, and most other experiments still work with a much smaller tweezer number. Furthermore, both AODs and SLMs lack the capability to produce light spots with sizes of several micrometers. As a result, a large, high-numerical-aperture microscope objective is required to focus these spots down to micrometer-scale sizes suitable for trapping a single atom (see Fig. \ref{comparing}(a) and (b)). This requirement not only introduces additional constraints on scalability related to the limited field of view of the microscope but also leads to reduced laser power due to optical losses and complicates system integration.

To overcome these limitations, optical metasurfaces have emerged as a powerful platform for generating optical tweezer arrays~ \cite{Zhan2024,chen2025multifunctional,Xu2025,holman2025}. A metasurface is a planar optical device composed of subwavelength nanostructures, known as meta-atoms, that can be individually engineered to modulate the phase, amplitude, and polarization of an incident light field~\cite{kildishev2013planar,Yu2014,jung2021metasurface}. Its operating principle is similar to that of an SLM; however, thanks to their nanoscale features, metasurfaces can integrate a much larger number of control pixels within a compact device. This large pixel number enables the generation of arrays comprising tens of thousands to millions of tweezers. Moreover, the subwavelength meta-atoms can be designed to produce sharp phase gradients, allowing pronounced light deflection and direct focusing down to micrometer-scale spot sizes~\cite{Khorasaninejad2016,Hsu2022,huang2023metasurface}. This dual capability, acting simultaneously as a large-scale tweezer generator and a focusing element, eliminates the need for an additional microscope objective (Fig. \ref{comparing}(c)). To demonstrate this capability, in this work, we designed and fabricated a metasurface that directly generates an array of $280 \times 280$ ($78,400$) optical tweezers with an intensity uniformity exceeding $90\%$.

\begin{figure}
    \includegraphics[width=\columnwidth]{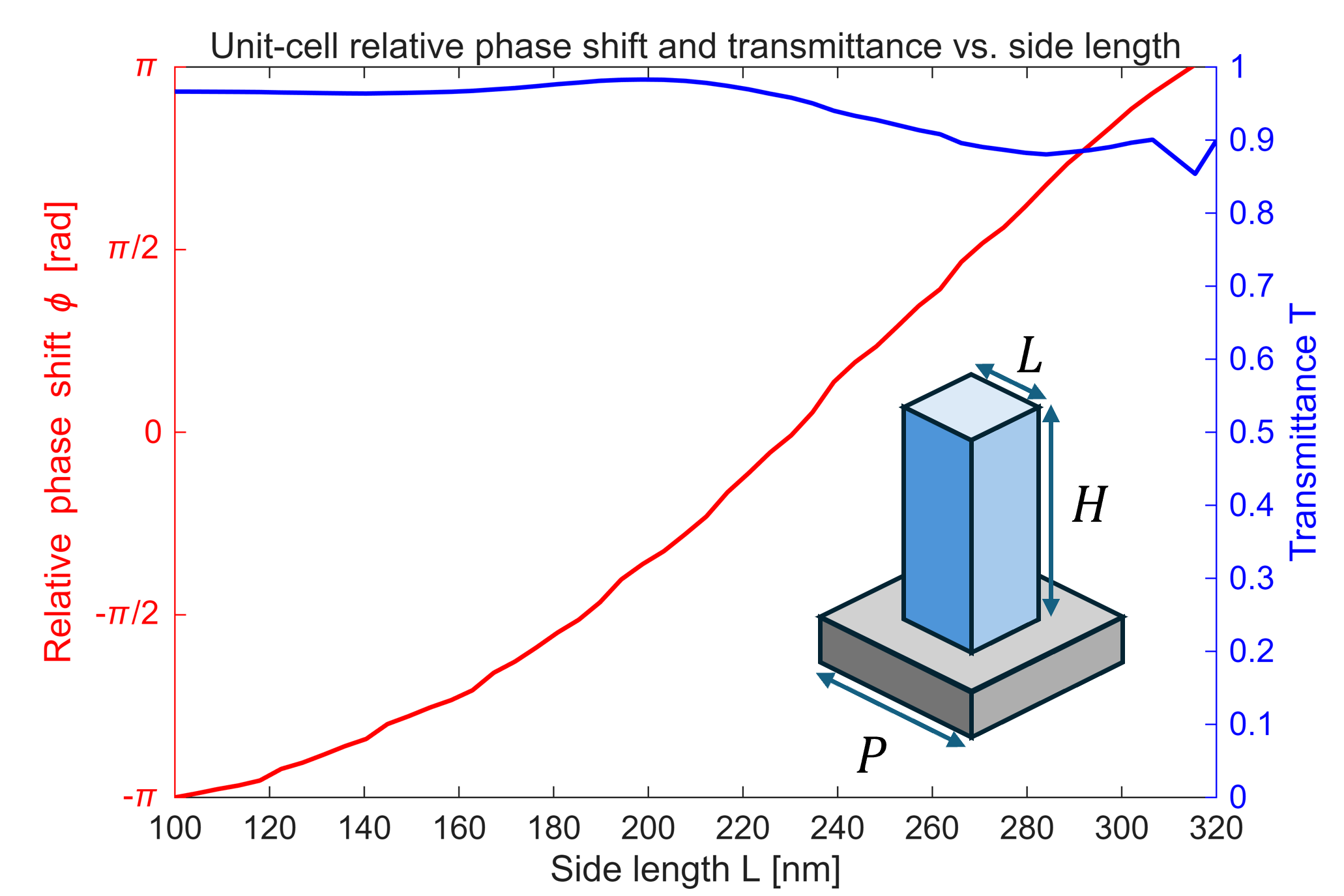}
    \caption{Plot of the relative phase shift $\phi$ (left axis) and transmittance $T$ (right axis) of a single meta-atom as a function of meta-atom size $L$, obtained via FDTD simulations.} 
    \label{fig_phase_trans}
\end{figure}

\textit{Meta-atom Parameters.} The metasurface is constructed using meta-atoms, square nanopillars made of silicon-doped silicon nitride. Each nanopillar has a side length $L$ and height $H$, and is arranged in a square lattice with period $P$, as shown in the inset of Fig.~\ref{fig_phase_trans}. The combination of the square pillar shape and square lattice geometry provides cross-sectional symmetry, enabling polarization-independent operation. The silicon doping ratio in silicon nitride is optimized to achieve a refractive index of $2.3$. Fig.~\ref{fig_phase_trans} presents the relative phase shift $\phi$ and transmittance $T$ of a single nanopillar unit at wavelength of 852~nm obtained through Finite-Difference Time-Domain (FDTD) simulations~\cite{sullivan2013electromagnetic}. The lattice constant $P$ and height $H$ are set to $430$~nm and $1150$~nm, respectively, while the side length $L$ varies from $100$~nm to $320$~nm. As shown in Fig.~\ref{fig_phase_trans}, the designed metasurface can provide relative phase shifts $\phi$ spanning the full range of $-\pi$ to $\pi$, with an average transmittance of over $85\%$. This plot also establishes a correspondence between the phase $\phi$ and the meta-atom size $L$ that will be needed later.

\textit{Design of Metasurface.} Here, we aim to fabricate a $5~$mm-diameter metasurface on a $10~\text{mm} \times 10~\text{mm}$ fused-silica substrate. This metasurface can generate a $280 \times 280$ array comprising $78,400$ optical tweezers. We employ the Weighted Gerchberg-Saxton (WGS) algorithm~\cite{DiLeonardo2007, kim2019gerchberg} to iteratively compute the phase profile at the metasurface required for generating the target optical tweezer array at the focal plane. Both forward and backward propagations are performed using the scalar angular spectrum method. The algorithm starts with an initial field, $E_{\text{inj}}$, defined by a uniform amplitude and a random phase across a $2.5$~mm-radius beam, sampled at the $430~\text{nm}$ unit cell period. In each iteration, the phase of $E_{\text{inj}}$ is updated based on the back-propagated field from the previous step, while its amplitude is held constant. After approximately $100$ iterations, the algorithm converges to a stable phase solution. Simulations show that the metasurface can generate optical tweezers with an Airy disk radius of $900~\text{nm}$ and an array intensity uniformity of $2.45\%$. 

The simulation of a $5~\text{mm}\times5~\text{mm}$ metasurface requires handling a large matrix of size $11{,}628 \times 11{,}628$, which poses a significant computational challenge. The memory demand of performing a standard Fast Fourier Transform (FFT) on such a matrix exceeds available resources. To address this issue, we adopted a block-wise FFT approach. Specifically, the matrix is partitioned into smaller blocks stored on a drive. Each block is loaded sequentially into memory for transformation, and the resulting output is written back to the drive before proceeding to the next block.

Based on the designed phase profile and the phase-size relationship established in Fig. \ref{fig_phase_trans}, the final meta-atom size distribution across the metasurface is determined.

\begin{figure*}
    \begin{center}  
    \includegraphics[width=2\columnwidth]{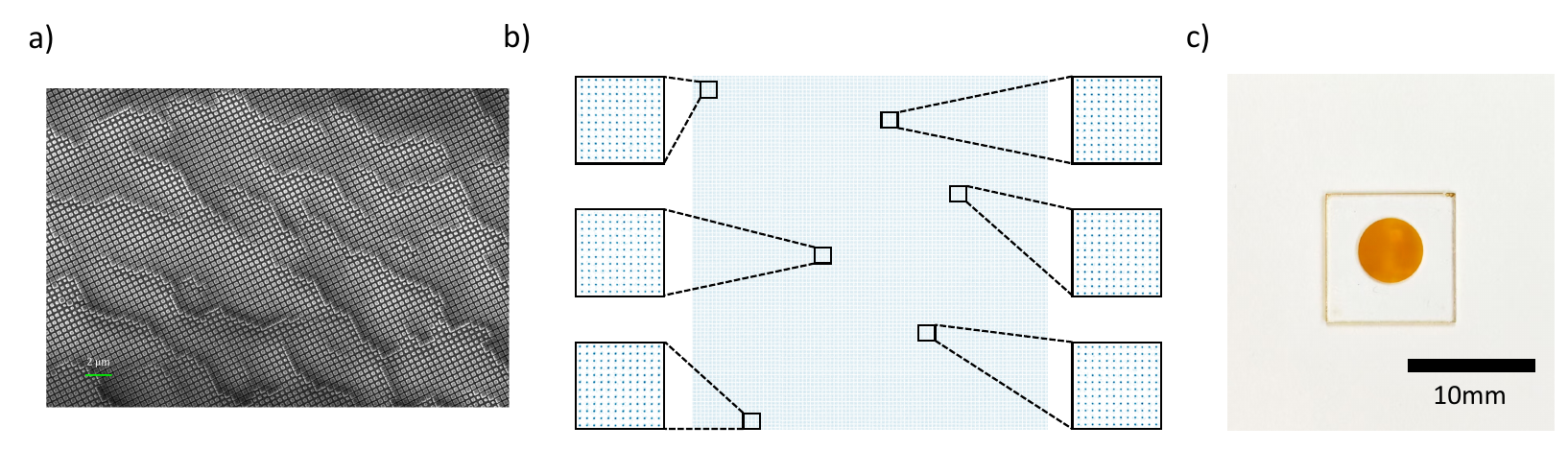}
\caption{Characterization of the metasurface and intensity distributions of the generated optical tweezers. (a) Scanning electron microscope image of the fabricated metasurface. (b) Image of the complete $280 \times 280$ tweezer array with zoomed-in views of representative areas. (c) An image of the metasurface employed in this study, which has a diameter of 5 mm, fabricated on a 10~mm$\times$10~mm  substrate} 
    \label{fig-imaging}
        \end{center}
\end{figure*}

\textit{Fabrication of Metasurface.} The metasurfaces are fabricated on fused-silica substrates using a CMOS-compatible process. The procedure begins with depositing a silicon nitride layer via plasma-enhanced chemical vapor deposition. A poly(methyl methacrylate) resist is then spin-coated and patterned by electron beam lithography. Subsequently, a chromium hard mask is formed through electron beam evaporation and lift-off. The pattern is transferred into the silicon nitride layer by inductively coupled plasma etching, followed by dry etching to remove the chromium mask. This well-established process robustly yields high-aspect-ratio silicon nitride nanopillars with aspect ratios exceeding $11:1$, characterized by sharp edges and smooth sidewalls. These features are essential for high-performance optical trapping.

\textit{Characterization of Metasurface.} Fig. \ref{fig-imaging}(a) shows a scanning electron microscope image of the fabricated metasurface. Each pillar functions as a super-atom, with approximately 1,300 such super-atoms corresponding to a single optical tweezer. To quantitatively evaluate its performance in generating the large-scale optical tweezer array, we constructed a precise optical characterization setup. A collimated $852$~nm laser beam with a waist diameter of $7.2$~mm was incident normally on the sample. The metasurface modulated the wavefront and focused it at a distance of $3.49$~mm, forming the desired array at the focal plane. The intensity distribution at the focal plane was then imaged by a high-numerical-aperture (NA = 0.65) microscope objective with a $0.45$~mm field of view and relayed to a high-resolution CCD camera for analysis. Fig. \ref{fig-imaging}(b) presents the intensity distribution from a representative section of the array, showing uniform and periodic features (see \cite{tweezerPicture} for full high-resolution images).

\begin{figure*}
    \begin{center}  
    \includegraphics[width=1.8\columnwidth]{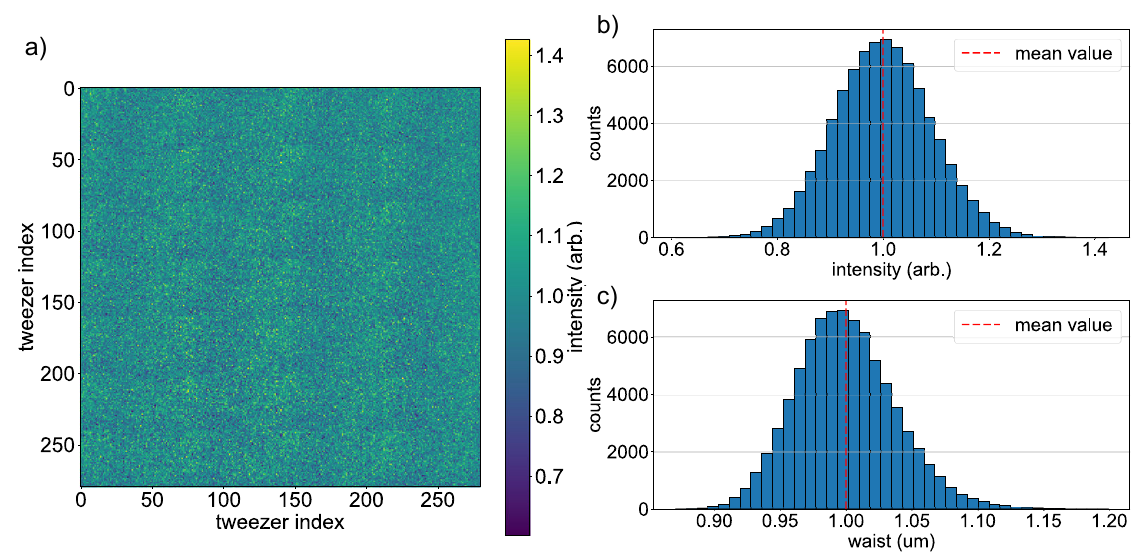}
    \caption{Characterization of the optical tweezer array generated by the metasurface: (a) Normalized intensity profile across the whole array. (b) Histogram of normalized intensities across the array with a standard deviation of 9.4\%. The intensity is obtained from the amplitude of the fitted Airy disk around each optical tweezer. (c) Histogram of the fitted airy disk waist.} 
    \label{fig4}
        \end{center}
\end{figure*}

Fig. \ref{fig4}~(a) shows the intensity distribution across the entire array, and quantitative analysis yields a 9.4\% standard deviation in the peak intensity variation in Fig. \ref{fig4}~(b). 
This measured nonuniformity significantly exceeds the simulated value of 2.45\%, likely due to fabrication imperfections in the meta-atoms, near-neighbor coupling effects and interference with the undiffracted light.

The measured center-to-center spot spacing is $4.258 \pm 0.044~ \rm\mu m$, consistent with the design target of $4.3~\rm\mu m$. 
Measurement of the light utilization efficiency shows that 67.5\% of the incident light on the metasurface is diffracted into the first diffraction order, a result consistently observed under both low- and high-power conditions, where the incident intensity was maintained at 15~W/cm², surpassing the typical efficiency of 45\% for SLM in high power~\cite{manetsch2025tweezer}. 
It is important to note that the efficiency calculation excludes the portion of light that does not illuminate the metasurface area. In practical experimental implementations, the incident beam size can be optimized to manage the inherent trade-off between the performance of the optical tweezer array and the overall power efficiency.
Meanwhile, 12.9\% of the light is modulated but does not contribute to the intended optical tweezers, instead appearing as unwanted background in the inter-trap regions. These two effects yield a final light utilization efficiency of 58.8\%.

To characterize the quality of each tweezer, we perform an analysis of the point spread function, which confirms that the profiles closely match the expected Airy disk shape, with over 99.7\% of the spots exhibiting Strehl ratios above 0.8. As shown in Fig. \ref{fig4}(c), the spots display near-diffraction-limited characteristics, with an average Airy disk radius of $ 1.017 \pm 0.038~\rm\mu m$. Compared to the designed numerical aperture (NA) of 0.58, this measured radius corresponds to an effective NA of 0.51 at the operating wavelength of 852 nm. We attribute this difference to the finite size of the incident light spot.

\textit{Discussion.} The scalability of atom arrays is constrained not only by the hardware for generating optical tweezers but also by the available laser power. Taking $^{87}$Rb atoms as an example, a trap depth of $\sim 0.5\ \text{mK}$ is generally required to achieve a high atom loading probability. To ensure both efficient trapping and long coherence time, optical tweezers are typically operated at a wavelength 
around $850$ nm. For tweezers with a numerical aperture of $0.5$, corresponding to an Airy disk radius of $\sim 1\ \rm\mu\text{m}$, the required optical power per tweezer is approximately $0.8$ mW. Currently available $850$ nm lasers can provide up to about $50$W of output power. After accounting for optical losses, including $10\%$ from free-space components such as isolators and $41.2\%$ due to the finite utilization efficiency of the metasurface, the usable power is considerably reduced, limiting the number of tweezers to roughly $33,000$ under ideal conditions. In this context, our demonstration of $78,400$ optical tweezers indicates that the scalability bottleneck for atom arrays has shifted from the constraints of tweezer generation hardware to the available laser power.

Metasurfaces as a powerful approach for optical tweezer generation have attracted growing interest from multiple research groups. For example, Ref. \cite{Zhan2024} utilized a dual-metasurface architecture that separates beam splitting and focusing to generate a $5 \times 5$ optical tweezer array, whereas our design employs only a single metasurface. 
Ref. \cite{chen2025multifunctional} used a multifunctional metalens to focus a trapping beam at 852~nm and collect single-photon fluorescence at 780~nm, thereby demonstrating the metasurface's remarkable flexibility, multifunctionality, and integration capabilities. However, this work did not address the scalability of such metasurfaces. 
Ref. \cite{Xu2025} demonstrated the vacuum-compatible generation of a $3\times 3$ array using a single metasurface; in contrast, we plan to position the metasurface outside the vacuum chamber. The most substantial scaling achievement to date is reported in an updated version of Ref. \cite{holman2025}, where approximately $360,000$ tweezers are produced using a $3.5\ \text{mm} \times 3.5\ \text{mm}$ metasurface. 
Although our metasurface design is similar, Ref. \cite{holman2025} uses about $300$ pixels per tweezer, approaching the simulation limit with uniformity large than $95\%$ and representing a landmark in pixel utilization efficiency. 
By comparison, we employ $1354$ pixels per tweezer to ensure robust beam quality. While it is technically feasible for us to reduce this number to their level and further increase the tweezer count, such optimization is unnecessary in our setup due to the geometric constraints. In practical experiments, placing the metasurface outside the vacuum imposes a minimum working distance defined by the vacumm cell dimensions, which requires a large size of metasurface and sufficient number of pixels. Given that the total number of traps is currently restricted by available laser power rather than pixel availability, the resulting pixel count per tweezer remains naturally well above the theoretical lower limit. 
Nonetheless, the tweezer numbers demonstrated in both works significantly exceed the current capabilities of AODs and SLMs, as well as the limitations imposed by available laser power. Thus, both results strongly validate the potential of metasurfaces for large-scale optical trapping. 

In summary, we have demonstrated the generation of a large-scale array of $78,400$ high-quality optical tweezers using a metasurface. This work establishes metasurfaces as a promising platform for scaling the number of qubits in atom-array systems to the order of $\sim 10,000$ in the near future.

\textit{Acknowledgement} 
This work is financially supported by National Natural Science Foundation of China (92576208), Tsinghua University Initiative Scientific Research Program, Beijing Science and Technology Planning Project, and Tsinghua University Dushi Program.

\bibliographystyle{iopart-num}
\bibliography{metasurface} 

\providecommand{\newblock}{}
\begin{thebibliography}{10}
\expandafter\ifx\csname url\endcsname\relax
  \def\url#1{{\tt #1}}\fi
\expandafter\ifx\csname urlprefix\endcsname\relax\def\urlprefix{URL }\fi
\providecommand{\eprint}[2][]{\url{#2}}

\bibitem{nielsen2010quantum}
Nielsen M~A and Chuang I~L 2010 {\em Quantum computation and quantum
  information\/} (Cambridge university press)

\bibitem{endres2016atom}
Endres M, Bernien H, Keesling A, Levine H, Anschuetz E~R, Krajenbrink A, Senko
  C, Vuletic V, Greiner M and Lukin M~D 2016 {\em Science\/} {\bf 354}
  1024--1027

\bibitem{barredo2018synthetic}
Barredo D, Lienhard V, De~Leseleuc S, Lahaye T and Browaeys A 2018 {\em
  Nature\/} {\bf 561} 79--82

\bibitem{ebadi2021quantum}
Ebadi S, Wang T~T, Levine H, Keesling A, Semeghini G, Omran A, Bluvstein D,
  Samajdar R, Pichler H, Ho W~W {\em et~al.\/} 2021 {\em Nature\/} {\bf 595}
  227--232

\bibitem{chiu2025continuous}
Chiu N~C, Trapp E~C, Guo J, Abobeih M~H, Stewart L~M, Hollerith S, Stroganov
  P~L, Kalinowski M, Geim A~A, Evered S~J {\em et~al.\/} 2025 {\em Nature\/}
  1--3

\bibitem{lin2025ai}
Lin R, Zhong H~S, Li Y, Zhao Z~R, Zheng L~T, Hu T~R, Wu H~M, Wu Z, Ma W~J, Gao
  Y {\em et~al.\/} 2025 {\em Physical Review Letters\/} {\bf 135} 060602

\bibitem{manetsch2025tweezer}
Manetsch H~J, Nomura G, Bataille E, Lv X, Leung K~H and Endres M 2025 {\em
  Nature\/}  1--3

\bibitem{bernien2017probing}
Bernien H, Schwartz S, Keesling A, Levine H, Omran A, Pichler H, Choi S, Zibrov
  A~S, Endres M, Greiner M {\em et~al.\/} 2017 {\em Nature\/} {\bf 551}
  579--584

\bibitem{Madjarov2020}
Madjarov I~S, Covey J~P, Shaw A~L, Choi J, Kale A, Cooper A, Pichler H,
  Schkolnik V, Williams J~R and Endres M 2020 {\em Nat. Phys.\/} {\bf 16} 857

\bibitem{cong2022hardware}
Cong I, Levine H, Keesling A, Bluvstein D, Wang S~T and Lukin M~D 2022 {\em
  Physical Review X\/} {\bf 12} 021049

\bibitem{Burgers2022}
Burgers A~P, Ma S, Saskin S, Wilson J, Alarcon M~A, Greene C~H and Thompson J~D
  2022 {\em PRX Quantum\/} {\bf 3} 020326

\bibitem{deist2022mid}
Deist E, Lu Y~H, Ho J, Pasha M~K, Zeiher J, Yan Z and Stamper-Kurn D~M 2022
  {\em Physical Review Letters\/} {\bf 129} 203602

\bibitem{bluvstein2022quantum}
Bluvstein D, Levine H, Semeghini G, Wang T~T, Ebadi S, Kalinowski M, Keesling
  A, Maskara N, Pichler H, Greiner M {\em et~al.\/} 2022 {\em Nature\/} {\bf
  604} 451--456

\bibitem{evered2023high}
Evered S~J, Bluvstein D, Kalinowski M, Ebadi S, Manovitz T, Zhou H, Li S~H,
  Geim A~A, Wang T~T, Maskara N {\em et~al.\/} 2023 {\em Nature\/} {\bf 622}
  268--272

\bibitem{Ma2023}
Ma S, Liu G, Peng P, Zhang B, Jandura S, Claes J, Burgers A~P, Pupillo G, Puri
  S and Thompson J~D 2023 {\em Nature\/} {\bf 622} 279

\bibitem{anand2024dual}
Anand S, Bradley C~E, White R, Ramesh V, Singh K and Bernien H 2024 {\em Nature
  Physics\/} {\bf 20} 1744--1750

\bibitem{zhang2024scaled}
Zhang B, Peng P, Paul A and Thompson J~D 2024 {\em Optica\/} {\bf 11} 227--233

\bibitem{finkelstein2024universal}
Finkelstein R, Tsai R~B~S, Sun X, Scholl P, Direkci S, Gefen T, Choi J, Shaw
  A~L and Endres M 2024 {\em Nature\/} {\bf 634} 321--327

\bibitem{muniz2025high}
Muniz J, Stone M, Stack D, Jaffe M, Kindem J, Wadleigh L, Zalys-Geller E, Zhang
  X, Chen C~A, Norcia M {\em et~al.\/} 2025 {\em PRX Quantum\/} {\bf 6} 020334

\bibitem{tsai2025benchmarking}
Tsai R~B~S, Sun X, Shaw A~L, Finkelstein R and Endres M 2025 {\em PRX
  Quantum\/} {\bf 6} 010331

\bibitem{bluvstein2025architectural}
Bluvstein D, Geim A~A, Li S~H, Evered S~J, Ataides J, Baranes G, Gu A, Manovitz
  T, Xu M, Kalinowski M {\em et~al.\/} 2025 {\em arXiv preprint
  arXiv:2506.20661\/}

\bibitem{zhang2025observation}
Zhang T, Wang H, Zhang W, Wang Y, Du A, Li Z, Wu Y, Li C, Hu J, Zhai H {\em
  et~al.\/} 2025 {\em Physical Review Letters\/} {\bf 135} 093403

\bibitem{Zhan2024}
Huang R, Zhou F, Li X, Xu P, Wang Y and Zhan M {\em Optics Express\/} {\bf 32}
  21293 ISSN 1094-4087

\bibitem{chen2025multifunctional}
Chen G~J, Zhao D, Wang Z~B, Li Z, Zhang J~Z, Chen L, Zhang Y~L, Xu X~B, Liu
  A~P, Dong C~H {\em et~al.\/} 2025 {\em Laser \& Photonics Reviews\/} {\bf 19}
  2401595

\bibitem{Xu2025}
Li D, Liao Q, Xu B, Zentgraf T, Castaneda E~N, Zhou Y, Qin K, Xu Z, Shen H and
  Huang L {\em {arXiv}\/}

\bibitem{holman2025}
Holman A, Xu Y, Sun X, Wu J, Wang M, Seo B, Yu N and Will S 2025 {\em arXiv\/}

\bibitem{kildishev2013planar}
Kildishev A~V, Boltasseva A and Shalaev V~M 2013 {\em Science\/} {\bf 339}
  1232009

\bibitem{Yu2014}
Yu N and Capasso F 2014 {\em Nat. Mater.\/} {\bf 13} 139

\bibitem{jung2021metasurface}
Jung C, Kim G, Jeong M, Jang J, Dong Z, Badloe T, Yang J~K and Rho J 2021 {\em
  Chemical Reviews\/} {\bf 121} 13013--13050

\bibitem{Khorasaninejad2016}
Khorasaninejad M, Chen W~T, Devlin R~C, Oh J, Zhu A~Y and Capasso F 2016 {\em
  Science\/} {\bf 352} 1190

\bibitem{Hsu2022}
Hsu T~W, Zhu W, Thiele T, Brown M~O, Papp S~B, Agrawal A and Regal C~A 2022
  {\em PRX Quantum\/} {\bf 3} 030316

\bibitem{huang2023metasurface}
Huang X, Yuan W, Holman A, Kwon M, Masson S~J, Gutierrez-Jauregui R,
  Asenjo-Garcia A, Will S and Yu N 2023 {\em Progress in Quantum Electronics\/}
  {\bf 89} 100470

\bibitem{sullivan2013electromagnetic}
Sullivan D~M 2013 {\em Electromagnetic simulation using the FDTD method\/}
  (John Wiley \& Sons)

\bibitem{DiLeonardo2007}
Di~Leonardo R, Ianni F and Ruocco G 2007 {\em Opt. Express\/} {\bf 15} 1913

\bibitem{kim2019gerchberg}
Kim H, Kim M, Lee W and Ahn J 2019 {\em Optics express\/} {\bf 27} 2184--2196

\bibitem{tweezerPicture}
High-resolution image of the complete $280 \times 280$ tweezer array
  \url{https://github.com/UltracoldTHU/MetasurfaceTweezerArray/raw/refs/heads/master/stitched_78400Tweezers_Generated_By_Metasurface.tif}
  accessed: 2025

\end{thebibliography}

\end{document}